# One-Dimensional Quantum Confinement Effect Modulated Thermoelectric Properties in InAs Nanowires

*Yuan Tian,[†,‡] Mohammed R. Sakr,[†,§] Jesse M. Kinder,[†] Dong Liang,[†] Michael J. MacDonald,[†] Richard L.J. Qiu,[†] Hong-Jun Gao,[*,‡] Xuan P.A. Gao [*,†]*

[†] Department of Physics, Case Western Reserve University, Cleveland, Ohio 44106, United States

[‡] Beijing National Laboratory for Condensed Matter Physics, Institute of Physics, Chinese Academy of Sciences, Beijing 100190, People's Republic of China

[§] Department of Physics, Faculty of Science, Alexandria University, Egypt

[*] Email: (X.P.A.G.) xuan.gao@case.edu; (H.J.G) hjgao@iphy.ac.cn

## Abstract

**We report electrical conductance and thermopower measurements on InAs nanowires synthesized by chemical vapor deposition. Gate modulation of the thermopower of individual InAs nanowires with diameter around 20nm is obtained over *T*=40 to 300K. At low temperatures (*T*< ~100K), oscillations in the thermopower and power factor concomitant with the stepwise conductance increases are observed as the gate voltage shifts the chemical potential of electrons in InAs nanowire through quasi-one-dimensional (1D) sub-bands. This work experimentally shows the possibility to modulate semiconductor**

**nanowire's thermoelectric properties through the peaked 1D electronic density of states in the diffusive transport regime, a long-sought goal in nanostructured thermoelectrics research. Moreover, we point out the importance of scattering (or disorder) induced energy level broadening in smearing out the 1D confinement enhanced thermoelectric power factor at practical temperatures (e.g. 300K).**

In the past five decades, much attention has been paid to thermoelectric materials due to their device applications in converting heat to electricity or vice versa.[1] This conversion efficiency is described by one fundamental parameter, the thermoelectric figure of merit, which is defined as $= S^2\sigma T/k$, where $S$ is the Seebeck coefficient or thermopower, $\sigma$ is the electrical conductivity, $T$ is the absolute temperature, and $k$ is the thermal conductivity. The difficulty in obtaining better $ZT$ values arises from the fact that S, $\sigma$, and $k$ are related to each other. Changing one alters the others simultaneously, which makes optimization extremely difficult. Since the first thermoelectric $Bi_2Te_3$ alloy has been found, the room-temperature $ZT$ of bulk semiconductors has increased only marginally, from about 0.6 to 1.[1,2] However, recent studies in nanostructured thermoelectric materials have led to significant advancements,[2,3] such as $Bi_2Te_3/Sb_2Te_3$ thin film superlattices,[4] PbSeTe/PbTe quantum dot superlattices[5] and nanocrystalline BiSbTe bulk alloys[6] with $ZT> 1$. There are two basic mechanisms by which $ZT$ is improved in low-dimensional systems: reduced lattice component of thermal conductivity with size effect, and increased power factor ($S^2\sigma$) by quantum confinement enhanced electronic density of states (DOS) in quasi-1D quantum wires/nanowires or quasi-zero-dimensional (0D) quantum dots or nano-particles. Most experimental works on the thermoelectric properties of nanostructures have found greatly reduced thermal conductivity in nanowires or nanoparticle composite materials. [6-9] However, the full potential of nanomaterials as high figure of merit thermoelectrics can only be achieved when the sharp (or singular) electronic DOS spectra in quantum confined low-dimensional (1D or 0D) systems are exploited, as first calculated by Dresselhaus in 1993 and elaborated further by others.[10-14] In this regard, combining the peaked 1D electronic DOS and drastically reduced thermal conductivity in semiconductor nanowires appears to be a promising route towards obtaining thermoelectrics with high figure of merit. Unfortunately, despite much progress in the study of thermoelectric properties of nanomaterials, [7-9, 15-17] there is no experimental work showing that the sharply peaked DOS in 1D or 0D materials can indeed be used to tailor the thermoelectric properties. In carbon nanotubes (CNTs) where the 1D electron confinement is strong, the thermopower was found to oscillate due to defects induced resonant electron scattering or the Coulomb blockade effect.[18, 19] In addition, due to the extremely high thermal conductivity of CNTs, it is not clear if CNTs can work as high ZT thermoelectric materials. In PbSe nanowires, the electrical gating method was used to modulate the Seebeck coefficient.[20] However, the generic behavior of the gate tuned Seebeck coefficient is consistent with bulk behavior and no sign of sharply peaked 1D DOS was observed. The underlying reason that it has been difficult to achieve 1D electronic confinement effect enhanced thermoelectric properties in semiconductor nanowires is tied to the stringent experimental requirements: first, nanowires are typically fairly large (10-100nm) in diameter such that low carrier concentration and low temperatures are generally required to reach the 1D regime for electron transport; second, even in the low carrier concentration/temperature regime, high quality material is required to avoid structural defect induced Coulomb blockade which can reduce the conductivity and dominate 1D electrical transport.

Here we report the thermoelectric properties investigated on single Indium Arsenide (InAs) nanowires. Thanks to the small effective electron mass ($m^*$=0.023$m_e$, with $m_e$ as the free electron mass), 1D sub-band splitting is large in InAs nanowires, making them a good candidate to study the 1D confinement effects on the thermoelectric properties at relatively high temperatures (e.g. above liquid nitrogen temperature and possibly even at room temperature). In a recent electrical transport study of InAs nanowire, the effect of 1D subband quantization was clearly seen as steps in conductance vs. gate voltage at moderate temperatures (100K) for nanowires with diameter 10-40nm.[21] Moreover, there are significant interest in the quantum transport effect [22, 23] and nanoelectronic device performance [24-29] of InAs nanowires. In this work, we demonstrate gate modulated thermopower together with the electrical conductance in InAs nanowires as the Fermi level sweeps through 1D electron subbands. Significantly, below temperature ~100K, thermopower exhibits oscillations concomitant with the stepwise conductance increase, as a result of the Fermi energy being tuned through the DOS peaks of 1D sub-bands. We further find that the gate modulated power factor oscillations correlate well with the corresponding behavior of 1D DOS spectrum. However, together with the thermal smearing of 1D energy levels, the scattering induced level broadening is found to severely suppress/smear out the oscillations in 1D thermoelectric transport properties, pointing to the imperative need for reduced carrier scattering to achieve quantum confinement enhanced thermoelectric figure of merit at practical temperatures. These findings provide the first evidence of 1D confinement induced thermopower modulation in semiconductor nanowires and serve to provide better insights into tailoring nanomaterials' thermoelectric properties.

InAs nanowires were grown by CVD method on silicon (100) substrates in 10% $H_2$/Ar mixture gas at 660°C with 20nm Au nanoparticles (Ted Pella, Inc.) as catalyst.[30] A Lindberg

Blue M tube furnace and 1inch diameter quartz tube were used for vapor transport synthesis of nanowires with accurate control of temperature and carrier gas flow rate. 99% InAs powder (Alfa Aesar) was ground as precursor and placed at the center of the furnace. The substrates were treated with poly-l-lysine (Sigma) for 8min30s and covered with 20nm Au catalyst for 5min successively. The pretreated substrates were then placed at a distance of 13.5cm from the source. Detailed growth conditions of InAs nanowires are as follows. At first, maximum flow rate of 100sccm (standard cubic centimeters per minute) $H_2$/Ar carrier gas was introduced after the system was pumped down to the base pressure of about 8mTorr. When the temperature of the furnace was increased to 660℃, the flow rate was decreased to 4sccm. The pressure was kept at 1.5Torr for 5min and then changed to 100mTorr for 3h. After that the furnace was cooled naturally down to room temperature and dark grey products composed of InAs nanowires were found on the substrates. These InAs nanowires were found to have zincblende crystal structure with <111> growth direction. [30]

As grown nanowires were suspended in ethanol by ultrasound sonication and then dropped onto a Si substrate with 600nm thick oxide. The Si substrate was degenerately doped and used as a backgate electrode. Photolithography and thermal evaporation of 80nm Ni were used to establish the electrode-patterns (Fig. 1a) designed for thermoelectric experiment. The sample was dipped in 0.5% hydrofluoric acid (HF) solution for 5s before evaporation to remove any native oxide layer and ensure Ohmic contacts. Low frequency lock-in techniques were used to measure the conductance and thermopower in a Quantum Design PPMS cryostat. In the thermopower experiment, a sinusoidal current at frequency ω was applied on the heater to generate a temperature gradient ΔT at frequency of 2ω. The resulting 2ω thermoelectric voltage $V_{th}$ along the nanowire was then detected by lock-in (SR830, Stanford Research). The temperature gradient ΔT was calibrated by using the four-wire resistances of the two electrode lines as thermometer #1 (TM1) and thermometer #1 (TM2) (supporting information). For the conductance measurement, a source-drain voltage of 100μV was applied between the electrodes (TM1 and TM2) at the ends of nananowire and the current was measured by a lockin (SR830). Multiple devices of InAs NWs with diameter about 25nm were measured and the manuscript presents data from the sample on which the most comprehensive data were collected (thermoelectric data on additional devices can be found in Fig. S4-7 in Supporting Information).

Fig. 1a shows a schematic and scanning electron microscope (SEM) micrograph of a typical device used in our study of thermoelectric properties. The line electrodes on top of InAs nanowire (zoomed in as the inset to Fig. 1c) serve also as thermometers. Another Ni electrode placed nearby, but not in contact with, the InAs nanowire serves as a line heater. A current $I_h$ applied to the heater causes a temperature gradient via Joule heating. The heating power (P) is calculated as $P = V_h^2/R_h$, in which $V_h = I_h R_h$ is the heater voltage, $R_h$ is the heater resistance. The temperature difference, ΔT, between the two thermometers is proportional to the heating power, which leads to a linear relation of thermoelectric voltage to P (Fig. 1b). The thermopower $S \equiv \frac{V_{th}}{\Delta T}$, was determined by the measured $V_{th}$ and the ΔT inferred from the resistance vs. P dependence of the two thermometers using their temperature dependent resistance (measured at equilibrium, P~0) as thermometry (Fig. S1, S2, Supporting Information). S is expected to be negative for n-type semiconductors, as we have confirmed for our n-type InAs nanowire. For the convenience of presentation, we show absolute value of S throughout this paper. Multiple devices of InAs NWs with diameter about 20nm were measured. The manuscript presents data from the sample (device #1) on which the most comprehensive data were collected. (Thermoelectric data on devices #2 and #3 can be found in Fig. S4-7 in Supporting Information.)

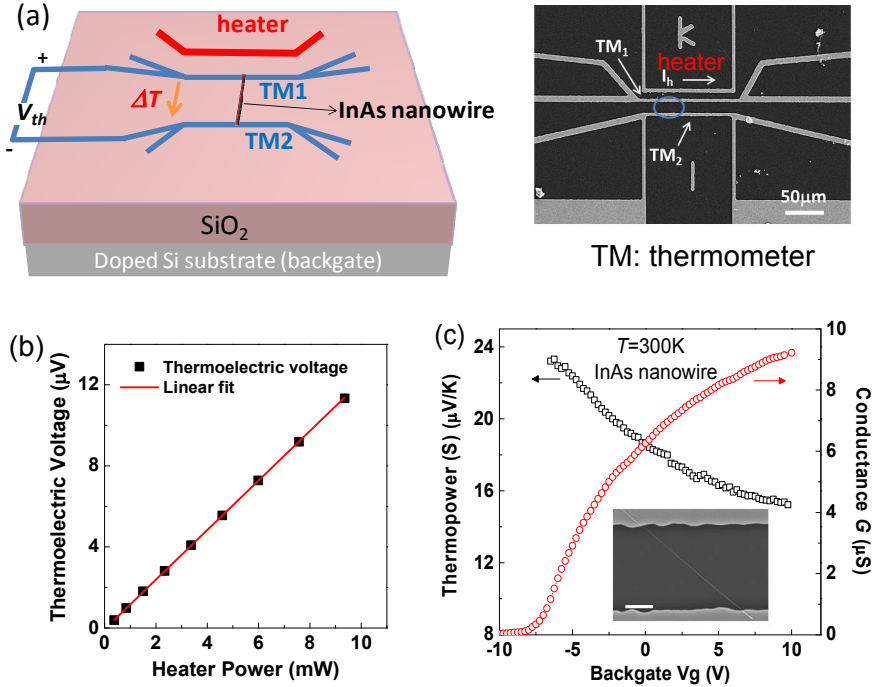

**Figure 1.** InAs nanowire thermoelectric transport device layout and the room temperature behavior. (a) A schematic and SEM image of nanowire thermoelectric device. Two line electrodes, $TM_1$ and $TM_2$, in contact with a single InAs nanowire, are used as thermometers to calibrate the temperature difference ΔT caused by the heating current $I_h$. (b) The Thermoelectric voltage ($V_{th}$) across the nanowire is proportional to the heating power ($P$) due to the proportion of $P$ to ΔT. (c) The gate dependence of InAs nanowire's conductance (red circles) and thermopower (black squares). The inset is a zoomed-in SEM image of the blue circle area in the SEM image of (a), showing two thermometer line electrodes connected by a single InAs nanowire. Scale bar: 2μm.

The room temperature ($T$=300K) gate voltage ($V_g$) dependence of the conductance $G$ and the thermopower $S$ of the InAs nanowire are shown in Fig. 1c. The $V_g$ dependence of $G$ is typical for a depletion mode n-type field effect transistor (FET). Using the diameter $d$=23nm and length of 9.5μm measured by SEM, an estimated peak field effect mobility about 2060cm$^2$/Vs is extracted by the trans-conductance value. In Fig. 1c, it is evident that as the InAs nanowire is gated towards depletion, the thermopower increases monotonically. (Similar behavior was observed in device #2 and #3, as shown in Fig.S4 and S6 in the Supporting Information). This behavior of increasing thermopower with decreased conductance or carrier density tuned by gate is consistent with the typical behavior of semiconductors [20, 31] and can be understood using Mott's formula for degenerate semiconductors [32]

$$S = -\frac{\pi^2}{3}\frac{k_B}{e}k_B T \frac{d\ln(\sigma)}{dE}\bigg|_{E=E_F} = -\frac{\pi^2}{3}\frac{k_B}{e}k_B T \frac{1}{G}\frac{dG}{dV_g}\frac{dV_g}{dE}\bigg|_{E=E_F} \quad (1)$$

where $E_F$ is the Fermi energy, $k_B$ is the Boltzmann constant and $e$ is electron charge. In Eq.1, $\frac{dG}{dV_g}\frac{dV_g}{dE_F}$ is positive and a monotonic function of $V_g$ (with increased $V_g$, the electron density increases so that both the Fermi energy and conductance increases). In such cases, the absolute value of thermopower $S$ is expected to increase monotonically as the gate tunes the nanowire towards the low carrier density/conductance regime. It should be noted that the measured thermopower values $S$ contains contributions from both the absolute values of the InAs nanowires $S_{InAs}$ and that of the Ni electrode $S_{Ni}$, i.e., $S = S_{Ni} - S_{InAs}$. But $S_{Ni}$ is expected to be insensitive to $V_g$ due to Ni being a good metal. Therefore the variation of $S$ can be purely attributed to that of InAs nanowire.

The $S$ vs. $V_g$ in Fig. 1c is in agreement with the general trend of bulk semiconductor's thermopower as a function of electron density (the electron density is proportional to the $V_g$ in our case).[20, 31] This is because the thermal broadening ($k_B T$) and scattering induced level broadening ($\delta E \sim h/\tau$) at 300K are too large to resolve the separation between quantized 1D subbands, where $k_B$ is the Boltmann constant, $h$ is the Planck's constant and $\tau$ is scattering time of electrons. When $k_B T$ and $\delta E$ are smaller than the energy separation between 1D subbands, the 1D subband quantization and the oscillating 1D DOS should play a dominant role in the thermoelectric transport properties as we demonstrate below. The conductance of the InAs nanowire as a function of the gate voltage $V_g$ is shown in Fig. 2a for temperatures from 40 to 300K. As can be seen, when $T$ decreases from 300K, the overall $G(V_g)$ curve shifts towards right and the trans-conductance at the low $G$ (electron density) increases, reflecting the carrier freeze out and increased electron mobility effects when semiconductors are cooled. What is more interesting is that below 100K, the curve develops some step-like features. These stepwise conductance increases are due to the population of 1D subbands and the Fermi level of the system lies between two adjacent 1D subbands at these steps.[21, 33] Fig. 2b presents a comparison between our experimental $G(V_g)$ data at 40K and numerically calculated curves for 1D subband occupation number $N_{1D\ SUBBANDS}$ vs. $V_g$ following Ref. 21.[34] The two fold degeneracy for subbands index higher than one was also included in the calculated $N_{1D\ SUBBANDS}$. We found that without scattering broadening, the thermal broadening itself produces steps (rises) flatter (sharper) than experimental data, consistent with Ref. 21. A close resemblance can be seen between the data and the theoretical curve (solid line) with both the thermal and scattering broadening are considered, corroborating the 1D subband filling interpretation of conductance plateaus. Note that the scattering broadening $\delta E$=28meV used in the calculation is in good agreement with the uncertainty principle estimation $h/\tau$ for the 40K mobility (~8000 cm$^2$/Vs) of our nanowire. As we discuss later, the scattering induced level broadening is actually the dictating factor in limiting the observation of 1D quantum confinement effect in thermopower and power factor at room temperature in our nanowires.

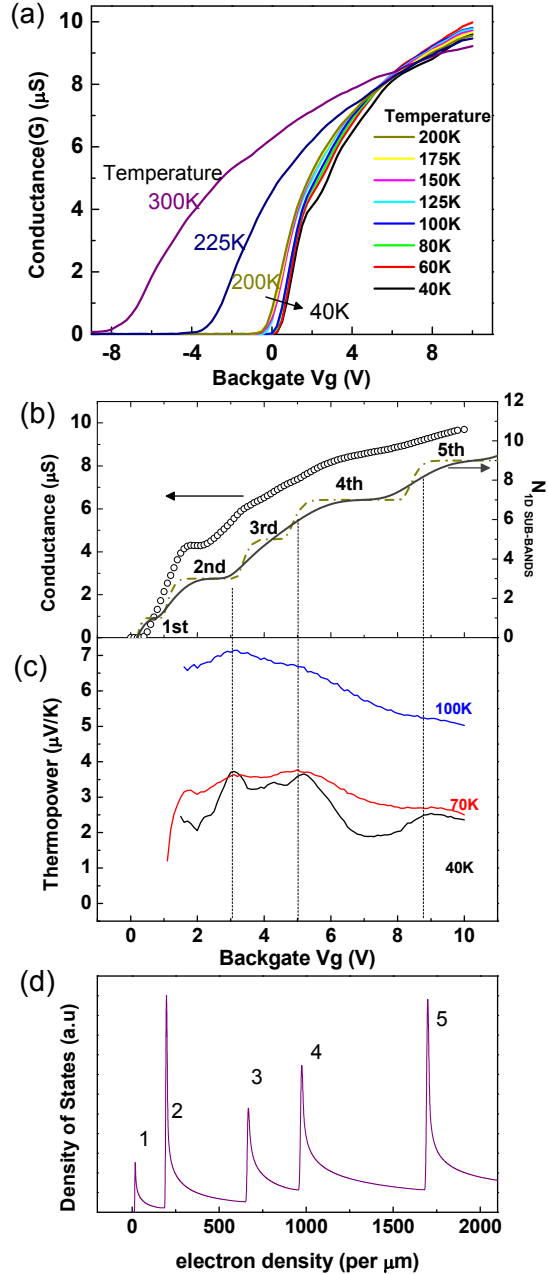

**Figure 2.** Gate tuned conductance and thermopower of InAs nanowire (diameter $d$=23nm) at different temperatures. (a) Temperature effect in the gate voltage dependent conductance $G$ from $T$=40K to 300K. Steps due to the electron population of individual 1D subbands as a function of $V_g$ are observed below 100K. (b) Comparing the measured 40K conductance vs. $V_g$ data (open circle) with the calculated 1D subband occupation with only thermal broadening (dash dotted

line) or both the thermal and scattering broadening (solid line) considered. (c) Gate modulation of thermopower $S$ at 100, 70 and 40K. The dashed vertical lines are a guide to the eye, highlighting the appearance of peak in $S(V_g)$ when a 1D subband starts to be filled. (d) Calculated density of states vs. 1D electron density in nanowire with the index of subbands marked.

Gate dependence data of thermopower $S$ at $T$=100, 70, and 40K are presented in Fig. 2c. As shown by Fig.2c, the $S(V_g)$ curve progressively deviates from the monotonically decreasing function as $T$ decreases. At $T$<100K, the gate voltage dependence of $S$ develops multiple oscillations. These oscillations are more discernible at lower $T$ and the peak positions coincide with the rises in the corresponding conductance vs. $V_g$ data when the next 1D subband starts to be filled (Fig.2d). (Similar effects were seen in devices #2 and #3, as shown in Fig. S5 and S7 in the Supporting Information.) Comparing Fig. 2b and c and the calculated 1D density of states in Fig.2d, we associate the peaks in thermopower data at $V_g$=3, 5, and 8.8V to be due to the filling of the third, fourth and fifth 1D subbands. Unfortunately, due to the slow thermal response of the device at low temperatures and the difficulty of using $2\omega$ thermopower measurement technique at frequency lower than 1Hz, we were unable to measure $S$ reliably down to the single subband regime where the sample impedance was high, although there might be some hint of the $2^{nd}$ subband filling effect in the lowest $V_g$ regime of our data.

More careful comparison of the conductance and thermopower data further shows a close connection between the electrical transport and thermopower data. In Fig.3, we plot $dG/GdV_g = dln(G)/dV_g$ at 40K together with the thermopower data. It can be seen that there is an excellent match of the peak/dip positions between the two traces: whenever $dG/GdV_g$ peaks/dips there is a peak or dip in the thermopower. This indicates Mott's formula works at least at the qualitative level for our data. Since conductivity $\sigma=ne^2\tau/m^*$, one sees from the Mott's formula $S\propto d(ln\sigma)/dE|_{EF}$ that both $S(V_g)$ and $dG/GdV_g$ should have characteristics reflecting the DOS, $dn/dE$, and $d\tau/dE$, or how the scattering time $\tau$ varies with energy. From the analysis in Fig. 3 and calculated 1D subband population in Fig. 2b, we believe that the oscillating 1D DOS is indeed mainly responsible for the oscillating $S$ and $dG/GdV_g$. We also show the original $G(V_g)$ curve in Fig. 3 for comparison. Examining the $G(V_g)$ curve carefully does reveal slope changes in connection with the thermopower oscillations. But since thermopower detects the derivative of energy dependent conductivity, characters of 1D DOS oscillations are more easily seen in $S(V_g)$ data.

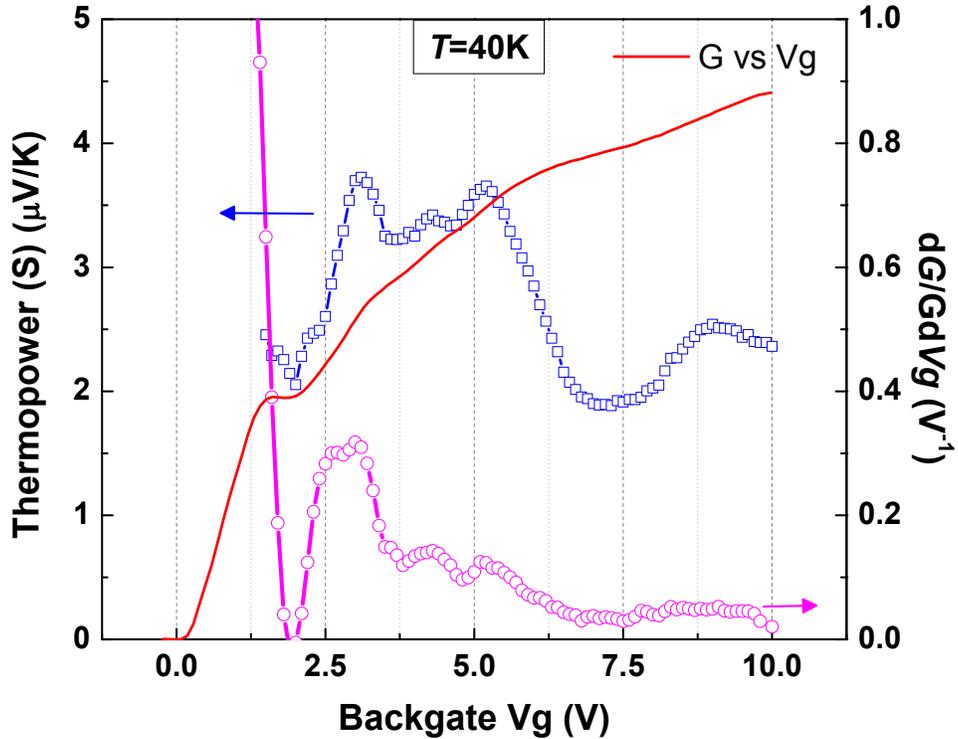

**Figure 3.** Gate dependence of thermopower, $S$ (blue), conductance $G$ (red) and $dG/GdV_g$ (magenta) at 40K, showing the correspondence between thermopower oscillations and the conductance steps as electrons populate 1D subbands in InAs nanowire.

This gate controlled subband filling gives us an opportunity to control and optimize the thermoelectric properties of nanomaterials. Here we consider the power factor $\sigma S^2$, a key parameter for the thermoelectrics since it is the numerator in the expression for *ZT*. Fig. 4 displays the gate modulated power factor at 300, 100, 70, and 40K. To make the discussion more meaningful/universal, we plot the $\sigma S^2$ as a function of electron density instead of gate voltage. As shown in Fig. 4a, power factor at 300K resembles that of a 3D bulk system, i.e. it is a smooth function peaked at some characteristic carrier density (~1600$e$/μm or 3.5×10$^{18}$/cm$^3$, marked by a star in Fig. 4a). Measurements on additional devices confirm that the power factor is always optimized at an electron concentration somewhat above 10$^{18}$/cm$^3$ at 300K for our InAs nanowires, although the absolute value of the maximal $\sigma S^2$ can vary from sample to sample depending on the mobility of particular nanowire (Fig. S4-7, Supporting Information). This optimal electron density for our nanowire is close to the theoretical value (~10$^{18}$/cm$^3$) for bulk InAs.[12] At 100K and below (Fig. 4b-d), the 1D quantum confinement effect induces oscillatory behavior due to the DOS of 1D subbands being resolved. (Since the thermopower *S* is roughly proportional to *T* as shown in Eq. 1, the absolute value of power factor decreases significantly as the temperature is reduced.) Apparently, the power factor is peaked when the 4$^{th}$ 1D subband starts to be populated at electron density ~1000 $e$/μm, a value slightly lower than the optimal *n* at 300K. Similar oscillatory behavior of power factor and the optimal electron density were observed in devices #2 and #3 (Fig. S5c and S7c, Supporting Information). We also point out that when the 1D subbands are resolved, the power factor exhibits much larger modulation (e.g. 500% change at 40 K and 70K vs. ~60% modulation at 300K) over the same carrier density range (Fig. S3, Supporting Information).

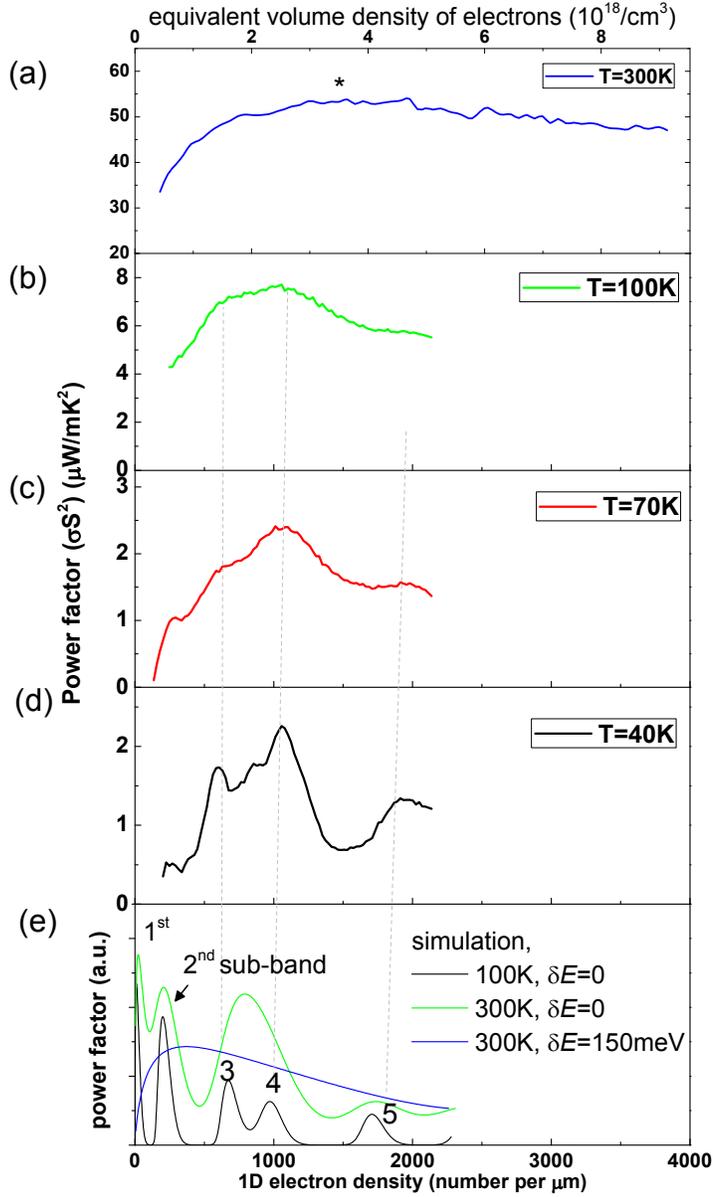

**Figure 4.** 1D quantum confinement effect in the power factor of InAs nanowire from 40 to 300K. (a-d) Gate modulated thermoelectric power factor $\sigma S^2$ at 300, 100, 70 and 40K plotted as $\sigma S^2$ vs. 1D density (bottom axis) or equivalent volume density (top axis) of electrons in InAs nanowire which has diameter $d$=23nm. The 1D subband effect is discernible at 100K and below. (e) Calculated power factor vs. electron density for circular InAs nanowire with $d$=23nm, under relaxation approximation and assuming $\tau \sim E^{-1/2}$. Different curves are results from calculations with or without scattering induced level broadening.

Before concluding the paper, we remark on additional insights obtained from our gate enabled power factor control via the 1D electron confinement in InAs nanowire. Originally, it was believed that the thermoelectric power factor is always maximized when the Fermi level is placed right at the position of singular DOS and smaller number of subbands is preferred.[10, 11] Our experiment indicates that the general idea of utilizing singular DOS to modulate the power factor in nanomaterials is effective. However, several important questions remain to be understood further such as where the optimal Fermi energy lies and if the system has to be in the single subband regime. Here, we found that although we cannot rule out possible upturn of power factor in the experimentally immeasurable one-subband regime, the power factor peak associated with the 3$^{rd}$ subband (and possibly 2$^{nd}$ subband) is clearly lower than the 4$^{th}$ one. Therefore, it is not clear if less subband filling will give rise to higher power factor. At low carrier density regime, there is also a practical issue of increased contact resistance that can limit the power factor. In addition, our result and analysis emphasize that making small sized material alone is not sufficient to induce low-dimensional confinement enhanced thermoelectric properties. One also has to reduce carrier scattering (improve the carrier mobility) such that the scattering broadening would not wash out the otherwise sharp DOS spectrum. For example, in Fig. 4e we present two simulated curves of $\sigma S^2$ vs. $n$ at 100 and 300K, assuming no scattering broadening ($\delta E$=0) and $\tau \sim E^{-1/2}$, a $\tau(E)$ dependence widely used for nanowires.[35] The calculation predicts power factor oscillations at locations close to the experimental data but the predicted oscillation amplitudes are much greater and increase as the number of subbands decreases. Moreover, the theoretical calculation with $\delta E$=0 exhibits clear oscillations even at 300K, in contrast to the experimental observation which is closer to the 3D behavior. We found that proper scattering broadening is needed in the calculation to improve the agreement. As shown by the blue line in Fig. 4e, adding 150meV scattering broadening (estimated from the room temperature mobility of nanowire) does wash out the 1D oscillatory features and restores the qualitative 3D behavior. Based on this observation, it appears that obtaining 1D confinement enhanced thermoelectric power factor at room temperature is possible in experimentally accessible parameter ranges for higher mobility InAs nanowires (e.g. 10-20nm wide InAs/InP core/shell nanowires with >$10^4 cm^2$/Vs mobility at room temperature[26]) such that both the thermal and scattering broadening effects are smaller than the subband separation energy. Note that the importance of scattering broadening of singular DOS spectrum was recently brought up from the theoretical perspective as well.[14]

In conclusion, we realized the gate controlled thermoelectric properties of InAs nanowires. Below 100K, gate voltage is shown to induce pronounced thermopower and power factor oscillations from the filling of 1D subbands for the first time in semiconductor nanowires in the diffusive transport regime. This work demonstrates the feasibility of experimentally optimizing 1D nanostructure's thermoelectric properties by tuning Fermi level across the peaks of 1D electronic density of states, a long-sought goal in nanostructured thermoelectrics research. Moreover, we point out the importance of reducing scattering induced level broadening in the pursuit of 1D quantum confinement enhanced thermoelectric properties in quasi-1D semiconductors at practical temperatures (e.g. 300K).

**ACKNOWLEDGEMENTS**

X. P.A. Gao acknowledges the NSF CAREER Award program (grant number DMR-1151534) and the donors of the American Chemical Society Petroleum Research Fund (grant number 48800-DNI10) for support of this research. H.-J. Gao acknowledges the NSF of China for funding support.

we convert the computed $N_{\text{1D SUB-BANDS}}$ vs. electron density $n$ curve into a plot of $N_{\text{1D SUB-BANDS}}$ vs. $V_g$ by using $n=C_gV_g/e$ where $C_g=\dfrac{L\ln\left(\frac{2h}{r}\right)}{2\pi\varepsilon\varepsilon_0}$ with an effective dielectric constant of 3.0. Such a reduced effective dielectric constant of $SiO_2$ is reasonable for backgated nanowire devices to take into account the vacuum gap and finite size effect (Wunnicke, O. Gate capacitance of back-gated nanowire field-effect transistors. *Appl Phys Lett* **2006**, *89*, 083102). Note that in the calculation of $N_{\text{1D SUB-BANDS}}$, we used 21nm as the nanowire diameter for the 1D subband energy level calculation, subtracting the native oxide thickness (~2nm) on the surface of nanowire.

(35) Seol, J.H.; Moore, A.L.; Saha, S.K.; Zhou, F.; Shi, L.; Ye, Q.; Scheffler, R.; Mingo, N.; Yamada, T. *Jour Appl Phys* **2007**, *101*, 023706.

**SYNOPSIS TOC**

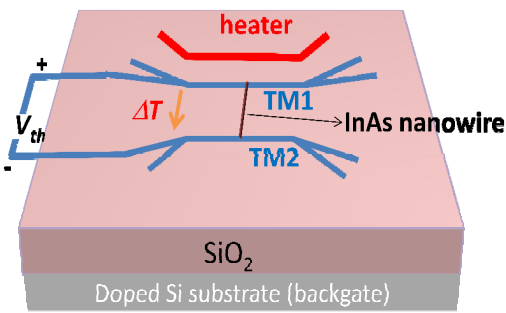
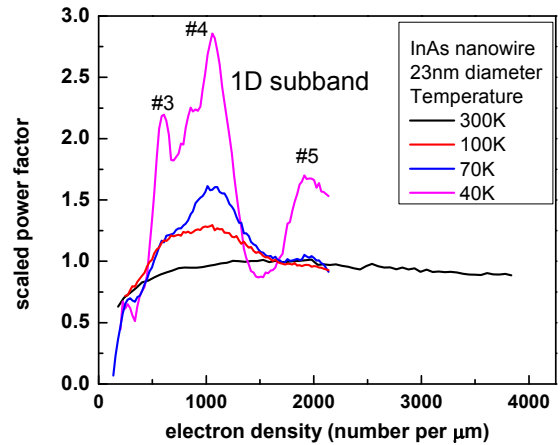